\documentclass[%
twocolumn,
superscriptaddress,
amsmath,amssymb,
aps,prx,
]{revtex4-2}
\usepackage{tabularx}
\usepackage{ltablex}
\usepackage{multirow}
\usepackage{amssymb,amsmath}
\usepackage{braket}
\usepackage[dvipsnames]{xcolor}
\usepackage{xcolor}
\usepackage{graphicx} 
\usepackage{dcolumn} 
\usepackage{bm} 

\usepackage{verbatim}
\usepackage{tikz}
\usepackage[compat=1.1.0]{tikz-feynman}
\tikzfeynmanset{warn luatex=false}
\usetikzlibrary{decorations.pathmorphing, decorations.markings, arrows.meta}

\usepackage[colorlinks=true,
            linkcolor=blue,
	        citecolor=blue,
	        urlcolor=blue]{hyperref}

\newcommand{\sissa}{Scuola Internazionale Superiore di Studi Avanzati (SISSA), I-34136 Trieste, Italy} 
\newcommand{\ictp}{The Abdus Salam International Centre for Theoretical Physics (ICTP), I-34151 Trieste, Italy}
\newcommand\mbf{\mathbf}

\begin{document}
\title{First- and Second-Order Electron Position Correlators Reveal Topological Phase Transitions
}

\author{Antimo Marrazzo}
\email{amarrazz@sissa.it}
\affiliation{\sissa}
\author{Christian Loer Llemit}
\affiliation{\sissa}
\affiliation{\ictp}
\author{Massimo Capone}
\affiliation{\sissa}
\date{\today}

\begin{abstract}
The theory of conventional critical phenomena is built around the concepts of fluctuations and correlators of local observables. 
Here, we show that  topological phase transitions can also be characterized through spatial and temporal two-point correlation functions associated with fluctuations of the electronic position operator. The key object is the antisymmetric part of the position-position correlation 
function, which yields the Chern number in the equal-time and equal-space limit.  When the system is translationally invariant in space or time (the latter case corresponding to an equilibrium condition),
these two-point correlators reduce to one-variable functions of the remaining variable. Remarkably, these functions are both reminiscent of critical behaviour, including divergent correlation times and lengths signalling the phase transitions. 
Away from these conditions, namely in non-periodic systems or genuinely dynamical processes, criticality is instead better captured by higher-order Chern-Chern correlators constructed from local or dynamical markers obtained as coincident-point limits of the position correlators. These higher-order Chern-Chern correlators can be seen as double Levi-Civita contractions of second-order position correlators, selecting the antisymmetric sector in each index pair.
We substantiate this framework through numerical simulations and analytic calculations. Our results establish a general correlator-based description of topological phase transitions, applicable to crystalline, noncrystalline, equilibrium, and dynamical systems.
\end{abstract}
 
\maketitle
\section{Introduction}
Topological phase transitions are often regarded as lying outside the conventional paradigm of critical phenomena, as distinct topological phases cannot, in general, be understood in terms of spontaneous symmetry breaking. However, the absence of a conventional Landau framework does not exclude an underlying description in terms of fluctuations and correlations~\cite{chen_prb_2017,favata_prl_2025}. Here we show that such a description emerges naturally from the fundamental objects of quantum geometry and topology.

Several key properties in condensed-matter physics, and electronic-structure properties in particular, are indeed often referred to as geometric and topological~\cite{vanderbilt_book_2018}. In many instances the Hamiltonian and its eigenstates depend on an external parameter that lives on a $D$-dimensional torus, where $D$ is the spatial dimension of the system. A most important example is the Bloch Hamiltonian $H(\mbf{k})$, which is defined on the differentiable Brillouin zone (BZ) torus. While the distance between two quantum states at different points of the torus is described by the quantum metric tensor~\cite{provost_cmp_1980}, their phase difference is related to the Berry curvature~\cite{vanderbilt_book_2018}. The two objects are, respectively, the real symmetric and imaginary antisymmetric parts of the same tensor, the quantum metric-curvature tensor (a.k.a. quantum geometric tensor or QGT)~\cite{berry_wil_WS_1989}:
\begin{eqnarray}
    \label{eq:f_def}
    \mathcal{F}_{\alpha\beta}(\mbf{k})&=&\sum_{n=1}^{J} \braket{\partial_{k_\alpha}u_{n\mbf{k}}|Q_{\mbf{k}}|\partial_{k_\beta}u_{n\mbf{k}}}\\
    \label{eq:g_def}
    g_{\alpha\beta}(\mbf{k}) &=& \text{Re}\,\mathcal{F}_{\alpha\beta}(\mbf{k})\\
    \label{eq:omega_def}
    \Omega_{\alpha\beta}(\mbf{k}) &=& -2\text{Im}\,\mathcal{F}_{\alpha\beta}(\mbf{k})
\end{eqnarray}
where $\alpha, \beta$ are Cartesian indices, $J$ is the number of occupied bands (i.e., number of electrons per unit cell), $Q_{\mbf{k}}$ is the projector on empty states and $\ket{u_{n\mbf{k}}}$ is the periodic part of the Bloch state $\ket{\psi_{n\mbf{k}}} = e^{i\mbf{k}\cdot\mbf{r}}\ket{u_{n\mbf{k}}}$.
Notably, the QGT is related to the second cumulant of the position operator, which in open boundary conditions (OBC) can be defined as~\cite{RS_prl_1999,swm_prb_2000}:
\begin{equation}
\braket{r_\alpha r_\beta}_c = \frac{1}{N}\braket{\Psi_0|r_\alpha r_\beta|\Psi_0}- \braket{\Psi_0|r_\alpha |\Psi_0}\braket{\Psi_0|r_\beta|\Psi_0}
\end{equation}
where $\Psi_0$ is the \emph{many-body} ground state and $N$ is the number of electrons. In passing, we note that many-body expressions of the localization tensor in periodic boundary 
conditions (PBC) have also been introduced~\cite{RS_prl_1999,swm_prb_2000}.
In the limit where $\Psi_0$ is a Slater determinant of Bloch orbitals, the localization tensor can be shown to reduce to the QGT~\cite{rr_jpcm_2002,sgiaro_prb_2001,rr_epjb_2011}:
\begin{equation}
\braket{r_\alpha r_\beta}_c = \frac{V_c}{J(2\pi)^D}\int_{BZ} d\mathbf{k}\,\mathcal{F}_{\alpha\beta}(\mbf{k})
\end{equation}
where $V_c$ is the unit cell $D$-dimensional volume. In particular, the trace of the quantum metric is related to the Resta-Sorella localization length~\cite{RS_prl_1999} and the gauge-invariant part of the Wannier function spread~\cite{mv_prb_1997}, while the imaginary part of the cumulant is related to the Chern number~\cite{rr_prl_2005}:
\begin{equation}
    \label{eq:loc_to_chern}
\text{Im}\braket{r_x r_y}_c = - \frac{1}{4\pi n_e}C,
\end{equation}
where $n_e=N/V$ is the number electron density (for a band insulator, $n_e =J/V_c$). For our purposes, we write the QGT in terms of the ground state projector $P$:
\begin{equation}
    \label{eq:cum_trace}
    \braket{r_\alpha r_\beta}_c = -\text{Tr}_{N}\left\{P[r_\alpha,P][r_\beta,P]\right\}
\end{equation}
where the trace is divided by the number of electrons~\footnote{The ground-state projector for a periodic system can be decomposed as $P= \frac{(2\pi)^D}{V_c}\int_{BZ}d\mbf{k}P_{\mbf{k}}$}. Hence, the modern theory of the insulating state relates two fundamental electronic properties\textemdash namely localization and non-trivial topology\textemdash to the fluctuations of the position operator. 

Here, we show that the cumulant of the position operator can be interpreted as the integrated equal-time or equal-space limits of suitable two-point time- and space-dependent correlation functions. These correlators display critical behaviour at topological phase transitions (TPTs) separating insulating phases with different Chern numbers. In equilibrium dynamics, temporal correlations depend only on the time difference, while in translationally invariant systems spatial correlations depend only on the relative displacement. As a result, they can be naturally analyzed in frequency and momentum space, respectively. The antisymmetric component of these correlators is directly sensitive to topology: its equal-time or equal-space limit is fixed by the Chern number, and its asymptotic sign also reflects the topological character of the phase. Close to a TPT, these one-variable correlators exhibit diverging correlation times and lengths. 

When the equilibrium and translation-invariance conditions are not met, namely in genuinely dynamical processes or non-periodic systems, criticality is instead more effectively captured by higher-order correlators between local Chern markers obtained from coincident-point limits of the underlying position correlators. 

\begin{figure*}
        \centering
        \includegraphics[width = \linewidth]{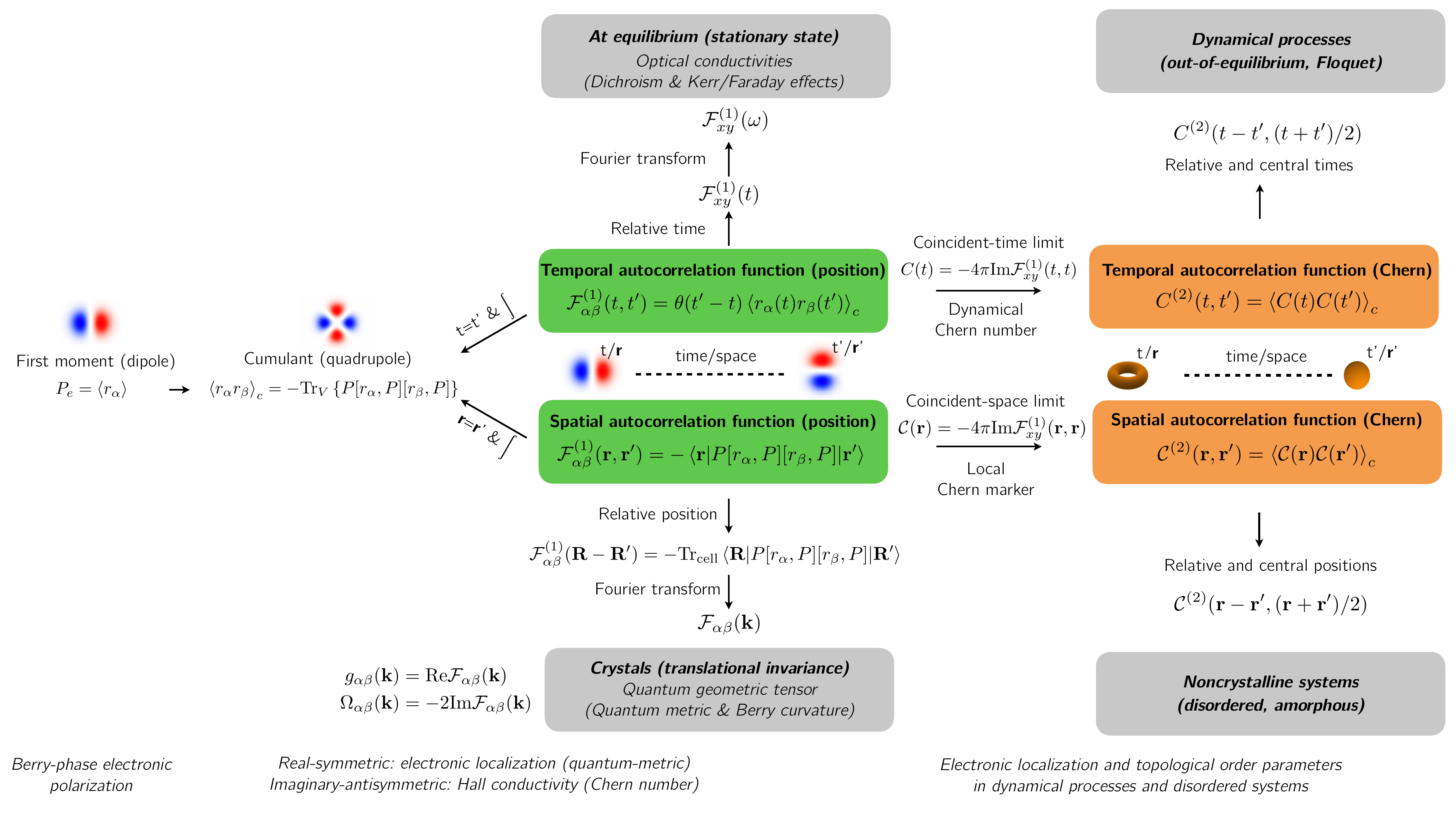}
        \caption{Overview of the theoretical framework. While Berry-phase polarization is related to the first moment, or dipole, of the position operator, the second cumulant is related to the quantum geometric tensor and captures both electron localization, through the quantum metric, and nontrivial topology, through the Chern number. This cumulant can be interpreted as the integrated equal-time or equal-space limit of temporal and spatial autocorrelation functions of the position operator ($\mathcal{F}_{\alpha\beta}^{(1)}$). In equilibrium or translationally invariant settings, these correlators reduce to one-variable functions that can be analysed in the corresponding dual spaces, namely frequency and momentum. 
        In the absence of these conditions, coincident-time or coincident-space limits of $\mathcal{F}^{(1)}_{\alpha\beta}$ define dynamical or local markers for the Chern number and localization. Connected autocorrelation functions of these markers can be obtained by double Levi-Civita contraction of the second-order correlators $\mathcal{F}^{(2)}_{\alpha\beta;\gamma\delta}$, with the contraction selecting the antisymmetric sector in each index pair.
        In particular, autocorrelation functions of the antisymmetric components provide a framework to describe two-dimensional topological phase transitions, with temporal correlations capturing dynamical processes and spatial correlations capturing disordered systems.}
        \label{fig:Overview}
    \end{figure*}

Our results therefore establish a general theoretical framework for describing TPTs in the language of critical phenomena, applicable to crystalline and noncrystalline systems, as well as to equilibrium and dynamical settings. 
Figure~\ref{fig:Overview} provides a schematic overview of this theoretical framework, that we propose as a conceptual roadmap for the manuscript highlighting the overarching logic underlying our approach.

\section{Spatial fluctuations}
First, we focus on spatial fluctuations and show that our theory provides a unified framework for several methods and results developed over recent years. In particular, we reinterpret local and nonlocal Chern markers in the language of correlation functions and critical phenomena. We start by expressing the trace in Eq.~\ref{eq:cum_trace} as a real-space integral of the local QGT marker $\mathcal{F}(\mathbf{r})$~\cite{marrazzo_prl_2019,resta_rnc_2018}:
\begin{equation}
    \label{eq:cum_int}
    \braket{r_\alpha r_\beta}_c = \frac{1}{N}\int d\mathbf{r}\,\mathcal{F}_{\alpha\beta}(\mathbf{r})
\end{equation}
where
\begin{equation}
    \label{eq:qgt_marker}
    \mathcal{F}_{\alpha\beta}(\mathbf{r}) = - \braket{\mathbf{r}|P[r_\alpha,P][r_\beta,P]|\mathbf{r}}.
\end{equation}
The imaginary and real parts of the QGT marker are related respectively to the local Chern marker~\cite{bianco_prb_2011,marrazzo_prb_2017,bau_prb_2024} and the localization (or quantum-metric) marker~\cite{marrazzo_prl_2019}: 
\begin{eqnarray}
    \mathcal{C}(\mathbf{r})&=&4\pi \text{Im}\braket{\mathbf{r}|P[r_\alpha,P][r_\beta,P]|\mathbf{r}},\\
    \mathcal{L}(\mathbf{r})&=& - \text{Re}\sum_{\alpha}\braket{\mathbf{r}|P[r_\alpha,P][r_\alpha,P]|\mathbf{r}}.
\end{eqnarray}
Their macroscopic averages describe respectively the local topology and the local electron localization, even in disordered and amorphous systems (see e.g.,~\cite{bianco_prb_2011,agarwala_prl_2017,marrazzo_prl_2019,marsal_pnas_2020, corbae_EPL_2023,bau_prb_2024,favata_prl_2025}). The local QGT marker (Eq.~\ref{eq:qgt_marker}) can be interpreted as the equal-space limit of the two-point spatial correlation function of the position operator, i.e.:

\begin{equation}
    \label{eq:qgt_corr}
    \mathcal{F}_{\alpha\beta}^{(1)}(\mathbf{r},\mathbf{r}') = - \braket{\mathbf{r}|P[r_\alpha,P][r_\beta,P]|\mathbf{r}'}.
\end{equation}
We note that similar two-point correlators were introduced in Refs.~\cite{chen_prb_2017,chen_njp_2019,molignini_scipost_2023,marsal_arxiv_2025} under the name of ``non-local'' markers and shown to exhibit critical behaviour at topological phase transitions.
 
The localization marker diverges at the TPT because the system approaches a gapless metallic state~\cite{marrazzo_prl_2019}. We note that, in band metals, the localization length generally diverges, while the $k$-space regular quantum metric defined as Eq.~\ref{eq:g_def} can remain finite~\cite{campos_prl_2007,rr_epjb_2011}. In semimetals, however, Eq.~\ref{eq:g_def} becomes singular as the direct gap closes at the Fermi level~\cite{nagaosa_prx_2020}.

\subsection{Crystalline systems} 
In the presence of crystalline order, translational symmetry enforces that the macroscopic average of the local Chern marker can be calculated over any primitive cell and is constant everywhere in the lattice. The non-local Chern marker can then be more conveniently written in terms of lattice vectors $\mathbf{R}$ as
\begin{equation}
    \label{eq:nlcm_cryst}
    \mathcal{C}(\mathbf{R}-\mathbf{R}') = 4\pi \text{Im}\,\text{Tr}_{\mathrm{cell}}\braket{\mathbf{R}|P[r_\alpha,P][r_\beta,P]|\mathbf{R}'},
\end{equation}
where the trace is taken over the occupied orbital inside one cell. Notably, the Chern number can be written as the equal-position autocorrelation function, i.e., $C=\mathcal{C}(\mathbf{R}=\mathbf{R}')$, and, due to translational symmetry, has the same value in every unit cell. In more general terms we introduce the position-position correlator
\begin{equation} 
    \mathcal{F}_{\alpha\beta}^{(1)}(\mathbf{R}-\mathbf{R}') = - \text{Tr}_{\mathrm{cell}}\braket{\mathbf{R}|P[r_\alpha,P][r_\beta,P]|\mathbf{R}'} 
\end{equation}
and take the discrete Fourier transform:
\begin{widetext}
\begin{equation}
     \sum_{\mbf{R}}e^{i\mathbf{k}\cdot\mathbf{R}}\mathcal{F}_{\alpha\beta}^{(1)}(\mathbf{R}) = \text{Tr}_{\mathrm{cell}}\{P(\mbf{k})\partial_{k_\alpha}P(\mathbf{k})\partial_{k_\beta}P(\mathbf{k})\}= \mathcal{F}_{\alpha\beta}^{(1)}(\mathbf{k}),
\end{equation} 
\end{widetext}
which is consistent with Eqs.~\ref{eq:f_def}-\ref{eq:omega_def} and where we used that $\left[r_{\alpha},P\right]=i\partial_{k_\alpha}P(\mathbf{k})$. Then it follows that the  antisymmetric part $\mathcal{F}_{xy}^{(1,\mathrm{A})}(\mathbf{R})=(\mathcal{F}_{xy}^{(1)}(\mathbf{R})-\mathcal{F}_{yx}^{(1)}(\mathbf{R}))/2$ is essentially the Fourier transform of the Berry curvature:
\begin{equation}
    \label{eq:f1a_from_berry}
\mathcal{F}_{xy}^{(1,\mathrm{A})}(\mathbf{R}) = -\frac{i A_c}{8\pi^2}\int_{BZ} d\mathbf{k}e^{-i\mathbf{k}\cdot \mathbf{R}}\Omega_{xy}(\mathbf{k}).
\end{equation}
We note that this quantity is related, but not equivalent, to the non-local Chern marker defined in Eq.~\ref{eq:nlcm_cryst}, since the operation of taking the imaginary part does not, in general, commute with the Fourier transform, whereas antisymmetrization does. 

Inversion and time-reversal symmetries impose distinct constraints on $\mathcal{F}_{xy}^{(1,\mathrm{A})}(\mathbf{R})$. In the presence of inversion symmetry, the Berry curvature is even in momentum space, implying that $\mathcal{F}_{xy}^{(1,\mathrm{A})}(\mathbf{R})$ is purely imaginary. Conversely, in the presence of time-reversal symmetry, the Berry curvature is odd, and $\mathcal{F}_{xy}^{(1,\mathrm{A})}(\mathbf{R})$ is therefore purely real.

\begin{figure}
    \includegraphics[width = 0.99\linewidth]{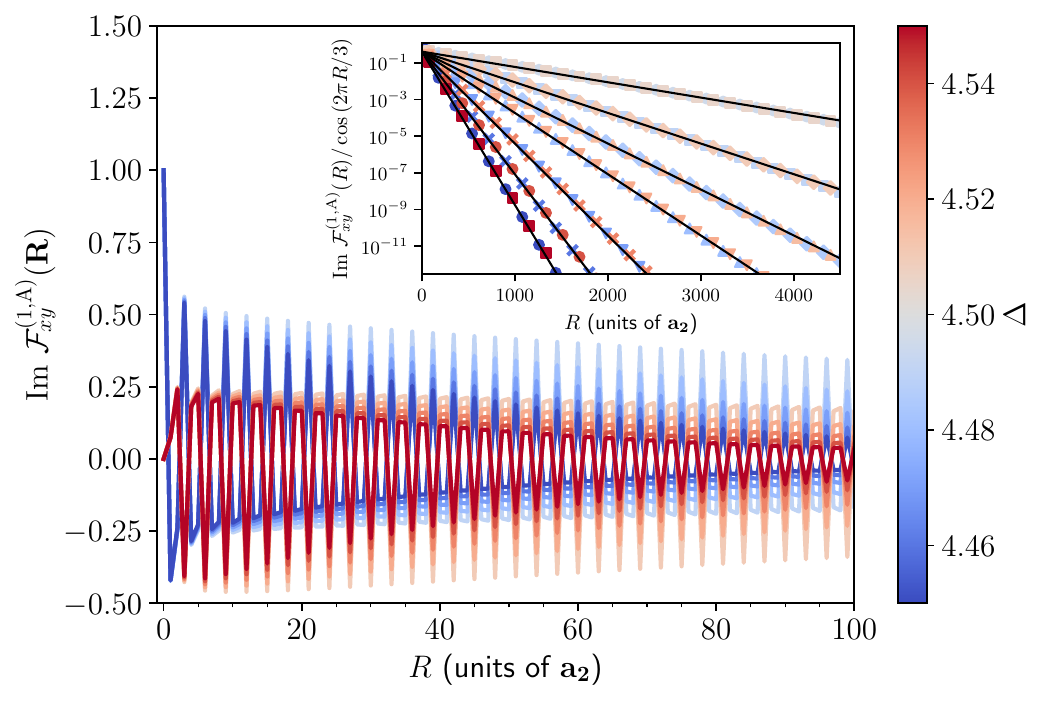}
        \caption{$\text{Im}\,\mathcal{F}_{xy}^{(1,\mathrm{A})}(\mathbf{R})$ in the pristine Haldane model for representative topological (blue) and trivial (red) phases, shown in units of $-A_c/4\pi$. At the origin, the correlator reproduces the Chern number, whereas at finite distances it decays to zero (see inset) according to the asymptotic form $\text{Im}\,\mathcal{F}_{xy}^{(1,\mathrm{A})}(\mathbf{r}) \propto \operatorname{sgn}(m) \cos(\mathbf{k}_{\sigma}\cdot\mathbf{r}) e^{-|m|r/v_F}$, where the Dirac fermion mass $m=\Delta-\Delta_c$ measures the distance from the critical point, with $\Delta_c=4.5$ for the parameters considered here. The oscillations have a period of three unit cells, set by the momentum of $K'$ at which the gap closes. }
        \label{fig:space}
    \end{figure}

In Fig.~\ref{fig:space}, we present the imaginary part of $\mathcal{F}_{xy}^{(1,\mathrm{A})}(\mathbf{R})$ for the Haldane model~\cite{haldane_1988}, obtained by numerically Fourier transforming the Berry curvature on a uniform $15{,}000\times15{,}000$ $\mathbf{k}$-point grid for representative topological and trivial phases (computational details in Appendix A). Appendices B and C include extended results for the Haldane and the Qi-Wu-Zhang (QWZ) models~\cite{qi2006topological}. We set the nearest-neighbour hopping $t_1=3$ and the complex next-nearest-neighbour hopping $t_2=1$ with phase $\phi=\pi/3$, and we study $\text{Im}\,\mathcal{F}_{xy}^{(1,\mathrm{A})}$ as a function of the on-site term $\Delta$, such that the TPT occurs for $\Delta_c=4.5$. We set $\mathbf{R}'=0$ and take $\mathbf{R}$ along the second lattice vector, $\mathbf{a}_2$. At $\mathbf{R}=0$, the correlator recovers the Chern number, taking the values $1$ and $0$ in the topological and trivial phases, respectively. For finite $\mathbf{R}$, it exhibits oscillations with a period of three unit cells while decaying to zero. Its asymptotic behaviour can be obtained analytically. As the system approaches the topological phase transition, the direct gap closes at the special point $K'$, and the corresponding low-energy contribution becomes increasingly singular. The long-distance behaviour can therefore be captured by a continuum Dirac Hamiltonian, as detailed in the SM. In this limit, the correlator decays as
\begin{equation}
    \label{eq:r_asymptotic}
\text{Im}\,\mathcal{F}_{xy}^{(1,\mathrm{A})}(\mathbf{r})\sim
\cos(\mathbf{k}_{\sigma}\cdot\mathbf{r})\,
\mathrm{sgn}(m)\,
e^{-\frac{|m|r}{v_F}},
\end{equation}
where $v_F$ is the Fermi velocity, $m=\Delta-\Delta_c$ is the Dirac mass, and $\mathbf{k}_{\sigma}$ is the special BZ point $K'$ where the gap closure occurs. While the oscillatory behaviour is set by the wave vector $K'$, the envelope decays exponentially with a correlation length $\xi=v_F/|m|$, consistent with other definitions of the correlation length based on finite-size scaling~\cite{caio_natphys_2019}. This behaviour is not specific to the particular Haldane transition considered here. As shown in Appendix B and C for both the Haldane and Qi-Wu-Zhang models, the same asymptotic form is recovered across transitions involving different Chern numbers and different changes of the topological invariant, demonstrating that the long-distance behaviour is controlled by the low-energy Dirac criticality rather than by the specific values of the Chern numbers.
\subsection{Disordered systems} 
In the pristine, translationally invariant system, the non-local Chern number captures the critical behaviour at the transition, whereas the local Chern number does not: translational invariance forces the local marker to be spatially uniform. Remarkably, the local Chern number has nevertheless been used to define a two-point connected Chern-Chern correlation function~\cite{favata_prl_2025}

\begin{equation}
    \label{eq:chern_corr}
    \mathcal{C}^{(2)}(\mathbf{r},\mathbf{r}') =  \braket{\mathcal{C}(\mathbf{r}) \mathcal{C}(\mathbf{r}')}_c.
\end{equation}
This can be generalized to a higher-order QGT-QGT correlator
\begin{equation}
    \label{eq:chern_corr}
    \mathcal{F}_{\alpha\beta;\gamma\delta}^{(2)}(\mathbf{r},\mathbf{r}') =  \braket{\mathcal{F}_{\alpha\beta}(\mathbf{r}) \mathcal{F}_{\gamma\delta}(\mathbf{r}')}_c,
\end{equation}
which is a rank-four tensor. Notably, the connected Chern-Chern correlator $ \mathcal{C}^{(2)}$ can be derived as the double Levi-Civita contraction of the connected rank-four QGT-QGT correlator $\mathcal{F}_{\alpha\beta;\gamma\delta}^{(2)}$, namely
\begin{equation}
    \mathcal{C}^{(2)}(\mathbf{r},\mathbf{r}') =  -4\pi^2  \varepsilon_{\alpha\beta}\varepsilon_{\delta\gamma}\mathcal{F}_{\alpha\beta;\delta\gamma}^{(2)}(\mathbf{r},\mathbf{r}') ,
\end{equation}
which selects the antisymmetric sector of each QGT factor independently.

 This higher-order Chern-Chern correlator has been shown to exhibit critical behaviour at TPTs in disordered systems~\cite{favata_prl_2025}. It was demonstrated that local topological markers behave as bona fide local order parameters in disordered two-dimensional topological insulators, with short-scale fluctuations that are removed by coarse graining away from the transition. At the disorder-driven topological phase transition, however, the connected Chern--Chern correlation function develops long-range tails and a divergent correlation length, revealing critical behaviour. Remarkably, the associated critical exponents appear universal within a given symmetry class and can distinguish different disorder-driven transitions, including topological Anderson transitions~\cite{favata_prl_2025}. 

The correlators $\mathcal{C}^{(2)}$ and $\mathcal{F}^{(2)}$ therefore probe fluctuations of the local topology\textemdash or, more generally, of the geometrical properties of the electronic structure\textemdash in real space. In a perfectly crystalline system, translational symmetry fixes the local markers to be spatially uniform, so that these fluctuations are suppressed and the connected correlator vanishes. In disordered or noncrystalline systems, instead, local variations of the QGT marker (i.e., of the coincident-space position correlator) become meaningful and can develop long-range correlations close to a topological phase transition.

From this perspective, $\mathcal{F}^{(1)}$ is the two-point correlation function associated with the second cumulant of the position operator, while $\mathcal{F}^{(2)}$ is a higher-order connected correlator built from products of local two-point responses. It is therefore related to higher-order fluctuations of the position operator, although it should not be identified directly with the fourth cumulant: rather, it represents the connected correlation of two second-cumulant, or QGT, densities.

\section{Temporal fluctuations}

Now we focus our attention on temporal fluctuations of the position operator, by introducing its time autocorrelation function\textemdash of which the cumulant in Eq.~\ref{eq:cum_trace} is the integrated equal-time limit:
\begin{eqnarray}
    \label{eq:f_1_ttp}
\mathcal{F}^{(1)}_{\alpha\beta}(t,t') &=& \theta(t'-t) \braket{r_\alpha(t) r_\beta(t')}_c \\
&=&   -\theta(t'-t)\text{Tr}_{N}\left\{P[r_\alpha(t),P][r_\beta(t'),P]\right\}
\end{eqnarray}
where the Heisenberg picture is understood and causality is enforced.
\subsection{Equilibrium correlations}
In the following, we consider systems at equilibrium, so the correlator depends only on time differences, i.e. $\mathcal{F}_{\alpha\beta}^{(1)}(t-t')$.  For a non-interacting system in PBCs  $\mathcal{F}_{\alpha\beta}^{(1)}(t)$  and its  Fourier transform read
\begin{widetext}
\begin{eqnarray}
    \label{eq:f_1_real_time_main}
\mathcal{F}^{(1)}_{\alpha\beta}(t) &=& \frac{\theta(t)V_c}{J(2\pi)^D}\int_{BZ}  d\mbf{k} \sum_{m\in\text{occ}}\sum_{n\in\text{unocc}} e^{-i\omega_{mn\mathbf{k}}t} \frac{\braket{u_{m\mathbf{k}}|v_{\alpha}|{u_{n\mathbf{k}}}}\braket{u_{n\mathbf{k}}|v_{\beta}|{u_{m\mathbf{k}}}}}{{\omega_{mn\mathbf{k}}}^2},\\
\mathcal{F}^{(1)}_{\alpha\beta}(\omega) &=& \frac{V_c}{J(2\pi)^D}\int_{BZ}  d\mbf{k} \sum_{m\in\text{occ}}\sum_{n\in\text{unocc}}  \frac{\braket{u_{m\mathbf{k}}|v_{\alpha}|{u_{n\mathbf{k}}}}\braket{u_{n\mathbf{k}}|v_{\beta}|{u_{m\mathbf{k}}}}}{{\omega_{mn\mathbf{k}}}^2}\cdot\frac{i}{\omega-\omega_{mn\mathbf{k}}+i\eta},
\label{eq:f_1_t_real}
\end{eqnarray}
\end{widetext}

where $v_\alpha$ is the velocity operator and $\eta$ is a small positive convergence parameter in the retarded form. 
This frequency-dependent QGT tensor has also been recently proposed in Refs.~\cite{queiroz_pnas_2025,rr_prb_2025} and, for 2D systems, can lead to a spectral decomposition of the Chern number~\cite{molignini_scipost_2023}:
\begin{eqnarray}
C &=& -4\pi n_e\text{Im}\,\mathcal{F}^{(1)}_{xy}(t=0^+) \\
&=& -\frac{2J}{A_c}\,\text{Im}\int_{-\infty}^{\infty} d\omega \mathcal{F}^{(1)}_{xy}(\omega) \in \mathbb{Z}
\label{eq:Chern_sum_rule}
\end{eqnarray}
where $A_c$ is the unit cell area. 

We now consider non-interacting band insulators across a TPT and monitor the behaviour of imaginary antisymmetric part $\text{Im}\,\mathcal{F}^{(1,\mathrm{A})}_{xy}$ in both the frequency and time domains. As a representative example and as we did previously for spatial fluctuations, we perform numerical simulations for the Haldane model in the topological and trivial phases, tuning the on-site term $\Delta$ across the transition. The results are shown in Fig.~\ref{fig:time}, while additional results for the Haldane and Qi-Wu-Zhang models are reported in Appendices B and C. In the frequency domain, the frequency-dependent QGT $\mathcal{F}_{\alpha\beta}^{(1)}(\omega)$ develops peaks at frequencies corresponding to direct interband transitions, $\omega=\omega_{mn\mathbf{k}}$. As the system approaches the TPT, the direct gap at the Dirac point $K'$ closes, and the corresponding low-energy contribution becomes increasingly singular. This gives rise to a strong enhancement of $\mathcal{F}_{xy}^{(1)}(\omega)$, which ultimately diverges at the critical point.

It is important to distinguish the gap-closing condition $\omega_{mn\mathbf{k}}=0$ from the resonant condition $\omega=\omega_{mn\mathbf{k}}$. In the latter case, the apparent singularity is regularized by the convergence factor entering the frequency-domain response, while in the former case the vanishing of the band gap produces a genuine critical divergence. We will return to this distinction below.
 \begin{figure*}[t]
        \centering
        \includegraphics[width = 0.48\linewidth]{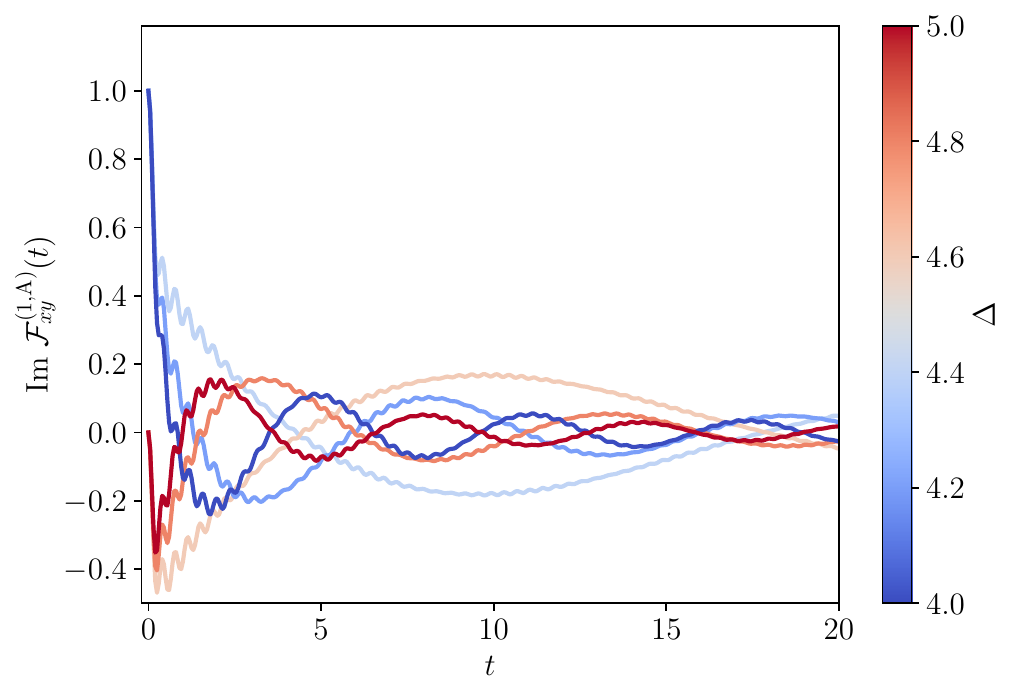}
       \quad
        \includegraphics[width = 0.48\linewidth]{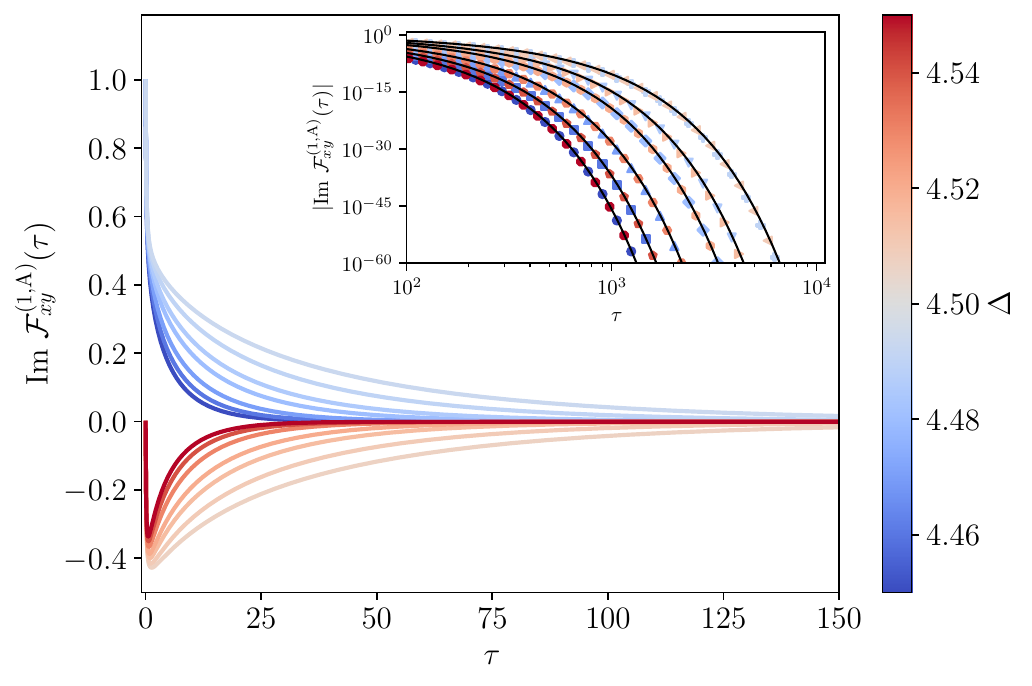}
        \caption{Imaginary part of $\mathcal{F}_{xy}^{(1,\mathrm{A})}$ for the Haldane model near the topological phase transition (TPT) for different values of the on-site term $\Delta$, shown in real time (left) and imaginary time (right). At $t=\tau=0$, $\text{Im}\,\mathcal{F}_{xy}^{(1,\mathrm{A})}$ equals the Chern number, in units of $-A_c/4\pi$. In real time, $\mathcal{F}_{xy}^{(1,\mathrm{A})}$ oscillates around zero, whereas in imaginary time it rapidly decays to zero. Close to the TPT, the imaginary-time decay acquires a power-law prefactor, $\mathcal{F}_{xy}^{(1,\mathrm{A})}(\tau)\sim \operatorname{sign}(m)\frac{e^{-2|m|\tau}}{2|m|\tau}$ where $m$ is the mass of the gapped Dirac cone at $K'$ as shown in the inset. Exactly at the critical point, $\mathcal{F}_{xy}^{(1,\mathrm{A})}$ is singular and is therefore not shown.}
        \label{fig:time}
    \end{figure*}

The same critical behaviour is reflected in the time domain, i.e. Eq.~\ref{eq:f_1_real_time}. In accordance with Eq.~\ref{eq:Chern_sum_rule}, at $t=\tau=0$  $\mathcal{F}_{xy}^{(1,\mathrm{A})}$ equals the Chern number (in units of $4\pi/A_c$). In real time, $\mathcal{F}_{xy}^{(1,\mathrm{A})}(t)$ displays oscillations around zero corresponding to the interband transition frequencies. In imaginary time, instead, $\mathcal{F}_{xy}^{(1,\mathrm{A})}(\tau)$ decays rapidly to zero away from the transition. Close to the TPT, this decay is controlled by the small Dirac mass $m$ of the gapped cone at $K'$, and acquires the asymptotic form (see Supplementary Material (SM)~\cite{SM})
\begin{equation}
        \label{eq:t_asymptotic}
\text{Im}\,\mathcal{F}_{xy}^{(1,\mathrm{A})}(\tau)
\sim
\operatorname{sgn}(m)\frac{e^{-2|m|\tau}}{2|m|\tau}.
\end{equation}
Thus, the imaginary-time correlator decays increasingly slowly as $|m|\to 0$, while exactly at the critical point $\mathcal{F}_{xy}^{(1,\mathrm{A})}(\tau)$ diverges.

The behaviour described so far is characteristic of a continuous phase transition: correlation functions develop increasingly long tails upon approaching the critical point, where the corresponding correlation scales diverge. In the present Dirac case, both the spatial correlation length, \(\xi\sim |m|^{-1}\), and the characteristic imaginary-time scale, \(\tau_c\sim |m|^{-1}\), diverge with exponent 1. This behaviour follows directly from the linear dispersion of the low-energy gapped Dirac cone and is therefore expected to be generic for topological transitions in this universality class. Higher-order band crossings can instead lead to different critical exponents and hence distinct universality classes, as demonstrated in Ref.~\cite{chen_prb_2017}.

Finally, we emphasize that $\mathcal{F}_{xy}^{(1)}(\omega)$ is directly connected to magneto-optical spectroscopy. In particular, its antisymmetric imaginary component is related to the frequency-dependent Hall conductivity $\sigma_{xy}(\omega)$ up to a simple frequency factor~\cite{rr_prb_2025}:
\begin{equation}
\text{Im}\,\sigma_{xy}^{(\mathrm{A})}(\omega) = \frac{e^2 J}{\hbar A_c}\omega \text{Im}\,\mathcal{F}_{xy}^{(1,\mathrm{A})}(\omega).
\end{equation}
Therefore, $\mathcal{F}_{xy}^{(1,\mathrm{A})}(\omega)$ contains the same spectral information that governs the transverse optical conductivity. In particular, $\sigma_{xy}(\omega)$ determines circular dichroism, namely the difference in absorption between left- and right-circularly polarized light, and enters the Faraday and Kerr rotation angles in transmission and reflection geometries, respectively~\cite{Tse_PRL_2010,Tse_PRB_2011,Pratama_PRB_2020,Sato_FPHYS_2022,Vignale_PRB_2023}. Hence, $\mathcal{F}_{xy}^{(1)}(\omega)$ can, in principle, be accessed through frequency-resolved magneto-optical probes, while the corresponding temporal correlator $\mathcal{F}_{xy}^{(1)}(t)$ can be obtained by Fourier transformation, including $\mathcal{F}_{xy}^{(1)}(\tau)$ in imaginary time by analytic continuation.

While here we focus on independent electrons, the relation between $\mathcal{F}_{xy}^{(1)}(\omega)$ and the optical Hall conductivity is completely general and follows from the Kubo formula, in which $\sigma_{xy}(\omega)$ is expressed in terms of the retarded current-current correlation function. This quantity can also be computed in interacting systems. When vertex corrections can be neglected, $\sigma_{xy}(\omega)$ is given by products of single-particle Green's functions after analytic continuation from Matsubara frequencies~\cite{kubo_jpsj_1957,Coleman_2015}. 
Thus $\mathcal{F}_{xy}^{(1)}(\omega)$ can be computed also in strongly correlated systems using Green's-function-based methods, such as dynamical mean-field theory (DMFT)~\cite{Georges1996}, in which vertex corrections can also be included~\cite{Lin2009,Toschi2012}. 
This perspective is particularly promising for investigating first-order TPTs driven by electron-electron interactions, as observed in topologically nontrivial two-orbital Hubbard models~\cite{sangio_prl_2015,Amaricci2016}.

While the short-time limit is determined by the Chern number, the decay of the real-time response may arise from two distinct mechanisms: dephasing and finite quasiparticle lifetimes. The corresponding correlation function is a superposition of contributions with different transition frequencies $\omega_{mn\mathbf{k}}$. Even without intrinsic damping, their phases progressively separate, leading to destructive interference and a decay of the total response. This dephasing does not imply the decay of the individual excitations and may give rise to partial revivals, particularly in finite systems with discrete spectra. Its characteristic timescale is approximately set by the spread of the relevant transition frequencies, $t_{\mathrm{deph}}\sim 1/\Delta\omega$. By contrast, a finite imaginary part of the retarded self-energy would give each excitation a linewidth $\Gamma_{mn\mathbf{k}}$, producing intrinsic real-time damping $e^{-\Gamma_{mn\mathbf{k}}t}$ through scattering or decay. The response may therefore contain both lifetime-induced exponential damping and a generally nonexponential dephasing envelope.

\subsection{Dynamical processes} 
Out of equilibrium, the temporal correlator $\mathcal{F}_{xy}^{(1)}$ cannot, in general, be expressed as a function of a single time or frequency variable. Following the same logic as in the spatial case, we introduce a higher-order correlation function by first considering the equal-time correlator $\mathrm{Im}\,\mathcal{F}_{xy}^{(1)}(t=t')$, which defines a time-dependent Chern number through Eq.~\ref{eq:loc_to_chern}. We then consider its connected autocorrelation function,
\begin{equation}
    \label{eq:chern_corr_time}
    \mathcal{C}^{(2)}(t,t') =
    \braket{\mathcal{C}(t)\mathcal{C}(t')}_c .
\end{equation}
Analogously to the position correlator, this quantity can be generalized to a higher-order QGT correlator
\begin{equation}
    \label{eq:F2_corr_time}
    \mathcal{F}_{\alpha\beta;\gamma\delta}^{(2)}(t,t') =
    \braket{\mathcal{F}_{\alpha\beta}(t)\mathcal{F}_{\gamma\delta}(t')}_c .
\end{equation}
As for spatial correlators, also $\mathcal{C}^{(2)}(t,t')$ can be seen as the double Levi-Civita contraction of the connected rank-four QGT-QGT correlator $\mathcal{F}_{\alpha\beta;\gamma\delta}^{(2)}(t,t')$.  These objects capture how topological fluctuations remain correlated between two different times during the dynamical evolution. While time-translation invariance is generally absent, one can still introduce central and relative time coordinates, $T_+=(t+t')/2$ and $T_-=t-t'$, and study the decay with relative time $T_-$ either at fixed central time $T_+$ or after averaging over $T_+$.

Finally, we note that dynamical Chern markers, which generalize the time-dependent Chern number to real space, have been introduced to study, for instance, Floquet dynamics~\cite{privitera_PRB_2016,privitera2017} and Kibble--Zurek behaviour in disordered Chern insulators~\cite{ul_prl_2020}. In the present framework, these can be derived from a generalized temporal \emph{and} spatial correlator of the position operator:
\begin{equation}
\mathcal{F}_{\alpha\beta}^{(1)}(\mathbf{r},\mathbf{r}',t,t') = - \braket{\mathbf{r}|P[r_\alpha(t),P][r_\beta(t'),P]|\mathbf{r}'},
\end{equation} 
of which the time-dependent Chern marker is the imaginary part of its equal-time and equal-space limit.

\section{Conclusions and outlook}
Our results recast topological phase transitions in the language of correlations and fluctuations: topological invariants arise as equal-space or equal-time limits of position correlators, while their spatial and temporal structure exposes the divergent scales associated with criticality.  Second-order rank-four correlators of electronic-position fluctuations  extend this picture when translational invariance or stationarity is lost: their double Levi-Civita contractions yield correlations between the corresponding local or dynamical topological markers, thereby capturing higher-order position fluctuations beyond the first-order description. Extending this approach to correlation-driven first-order topological transitions~\cite{Amaricci2016}, and reconstructing the same correlators from frequency-resolved optical Hall measurements—including circular dichroism and Kerr or Faraday spectroscopy—are natural next steps. More broadly, this framework offers a route towards describing topological phases and their transitions\textemdash from theory and simulation to experiment\textemdash within the same language of correlations and fluctuations.

\section*{Data Availability}
    The data that support the findings of this article are available upon reasonable request and will be publicly available on the Materials Cloud Archive~\cite{materialscloud_data} at the time of publication.

\section*{Acknowledgments}
A.M. acknowledges useful discussions with R. Resta.

\section*{Author contributions}
A.M. initiated the project and, together with M.C., developed its conceptual framework. A.M. carried out the analytical calculations and prepared the initial draft with input from C.L.L. and M.C. C.L.L. performed the numerical simulations and generated the corresponding figures under the supervision of A.M. and M.C. All authors analysed the results, discussed the findings, and contributed to revising and finalizing the manuscript.

\section*{Appendix A: Computational details}
For spatial correlators computed through Eq.~\ref{eq:f1a_from_berry}, the Berry curvature was obtained on a discrete $k$-point mesh of size $15{,}000\times15{,}000$ using the Fukui-Hatsugai-Suzuki algorithm~\cite{fukui2005chern} and then a 2D numerical FFT was used.

For temporal correlators in two-band models at half filling the discretized expression for the time-dependent QGT  at equilibrium simplifies to

    \begin{eqnarray}
    \label{eq:f_1_real_time}
   \mathcal{F}^{(1,\mathrm{A})}_{\alpha\beta}(t) &=& \frac{i\theta(t)}{N_k} \sum_{k}  e^{-i\omega_{01\mathbf{k}}t} \text{Im}\frac{\braket{u_{0\mathbf{k}}|v_{\alpha}|{u_{1\mathbf{k}}}}\braket{u_{1\mathbf{k}}|v_{\beta}|{u_{0\mathbf{k}}}}}{{\omega_{01\mathbf{k}}}^2}\\
   &=& -\frac{i\theta(t)}{2N_k} \sum_{k}  e^{-i\omega_{01\mathbf{k}}t} \Omega(\mathbf{k})
    \end{eqnarray}    

    where $N_k$ is the number of k-points sampled in the Brillouin zone, $\omega_{01\mathbf{k}}$ is the energy difference between the normalized unoccupied state $\ket{u_{1\mathbf{k}}}$ and occupied state $\ket{u_{0\mathbf{k}}}$ at momentum $\mathbf{k}$, and $v_{\alpha}$ is the velocity operator in the $\alpha$ direction. $ \mathcal{F}^{(1,\mathrm{A})}_{\alpha\beta}(t)$ was calculated through the sum over states (Eq.~\ref{eq:f_1_real_time}), the Brillouin zone was discretized with a converged uniform dense 2D k-point grid of size $3600 \times 3600$.

\section*{Appendix B: Extended numerical results for the Haldane Model}
\begin{figure*}[tbph]
        \centering
        \includegraphics[width = 0.48\linewidth]{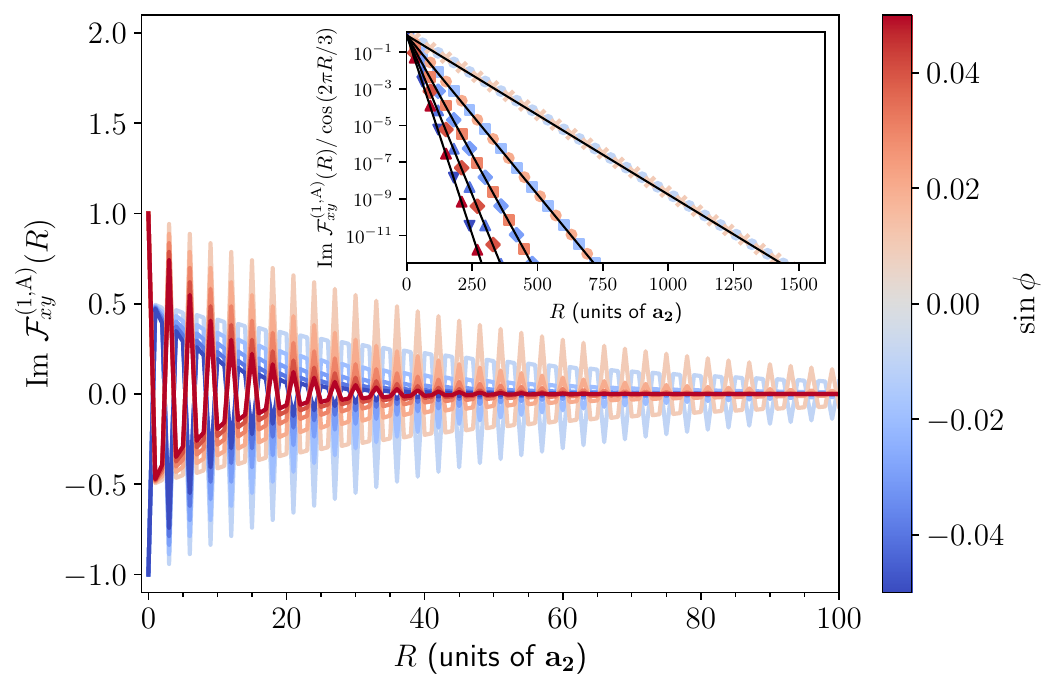}
        \quad
        \includegraphics[width = 0.48\linewidth]{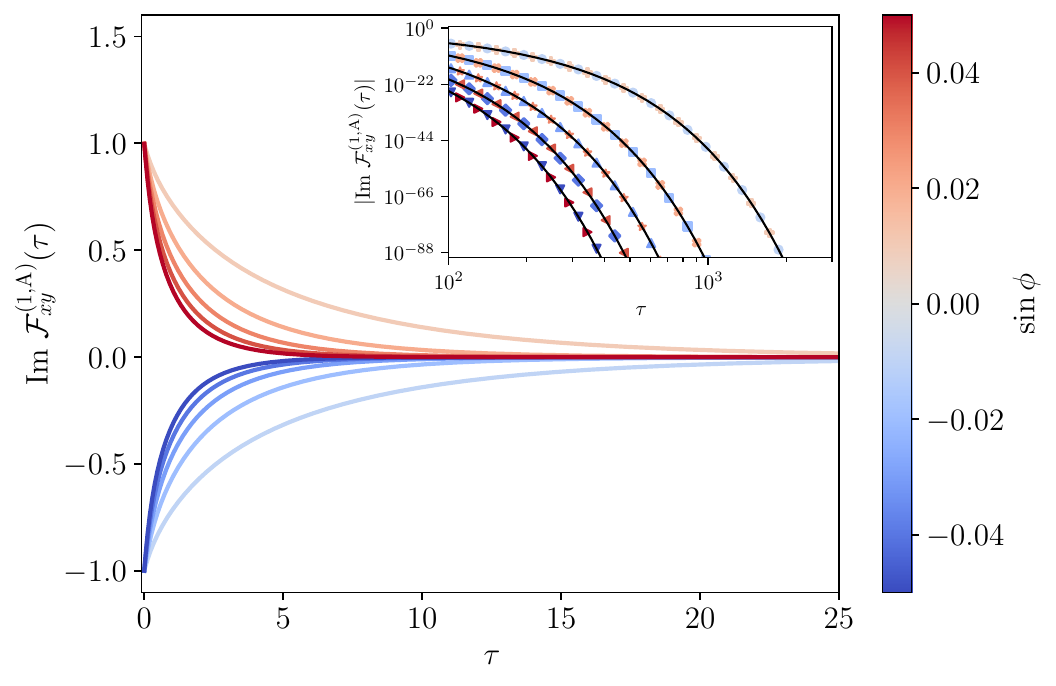}
        \caption{Numerical results for the pristine Haldane model across the topological transitions from $C=-1$ to $C=1$, showing $\text{Im}\,\mathcal{F}_{xy}^{(1,\mathrm{A})}({R})$ (left-hand panel) and $\text{Im}\,\mathcal{F}_{xy}^{(1,\mathrm{A})}(\tau)$ (right-hand panels). The insets demonstrate the agreement between the numerical results and the analytical predictions for the asymptotic behavior.}
        \label{fig:haldane_nonlocal}
    \end{figure*}

Here we present additional numerical results for the Haldane model when the Chern number directly changes from -1 to 1. This happens when the parameters are set to $t_1=3$, $t_2=1$, and $\Delta=0$, while $\phi$ is varied, corresponding to graphene for $\phi=0$. We report $\mathcal{F}^{(1,\mathrm{A})}(\mathbf{R})$ and $\mathcal{F}_{xy}^{(1,\mathrm{A})}(\tau)$ respectively in the left-hand and right-hand panels of Fig.~\ref{fig:haldane_nonlocal}. At the origin, the correlator reproduces the Chern number and decays to zero (see inset) according to Eq.~\ref{eq:r_asymptotic} where the Dirac fermion mass $m=\Delta-\Delta_c$ measures the distance from the critical point. The oscillations have a period of three unit cells due to $C_3$ symmetry. The gap closes at $\phi_c=0$ set by momentum $\mathbf{K}'$ and $\mathbf{K}$. Decay for $\mathcal{F}_{xy}^{(1,\mathrm{A})}$ in imaginary time follows Eq.~\ref{eq:t_asymptotic}. 

\section*{Appendix C: Numerical results for the QWZ Model}
\begin{figure*}[!htb]
    \centering  \includegraphics[width=0.4\linewidth]{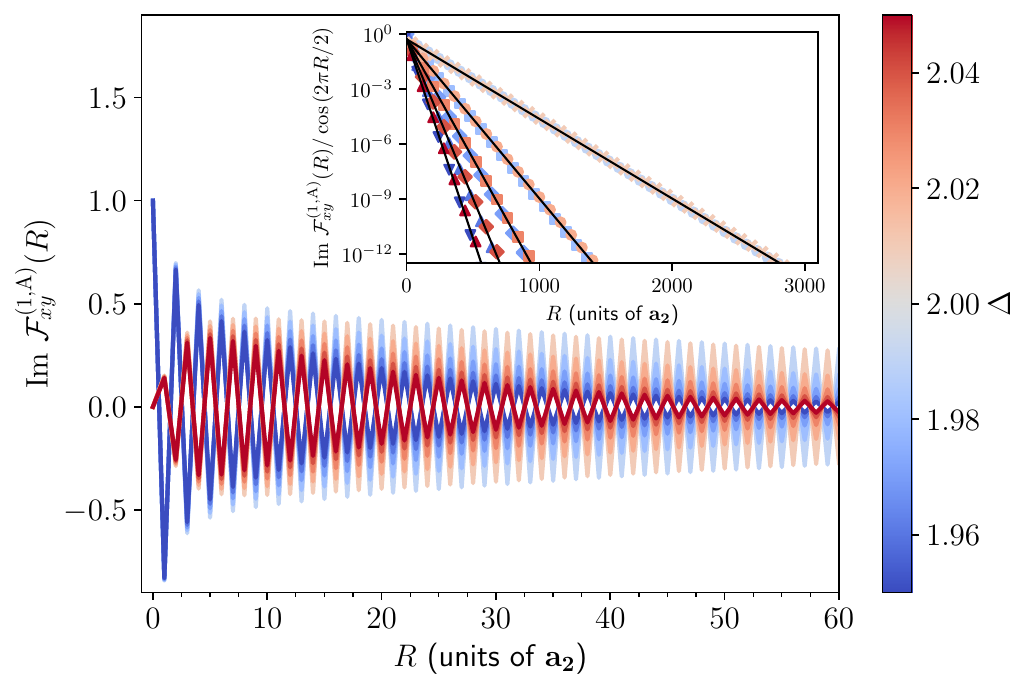} 
    \qquad \quad  \includegraphics[width=0.4\linewidth]{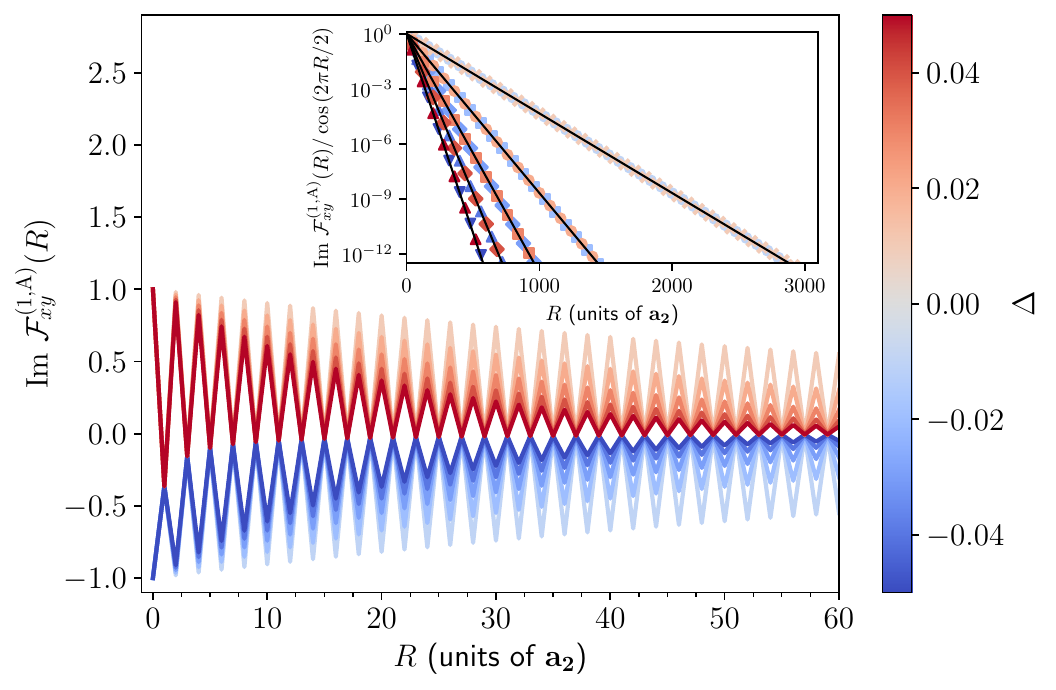}
     \qquad \quad  \includegraphics[width=0.4\linewidth]{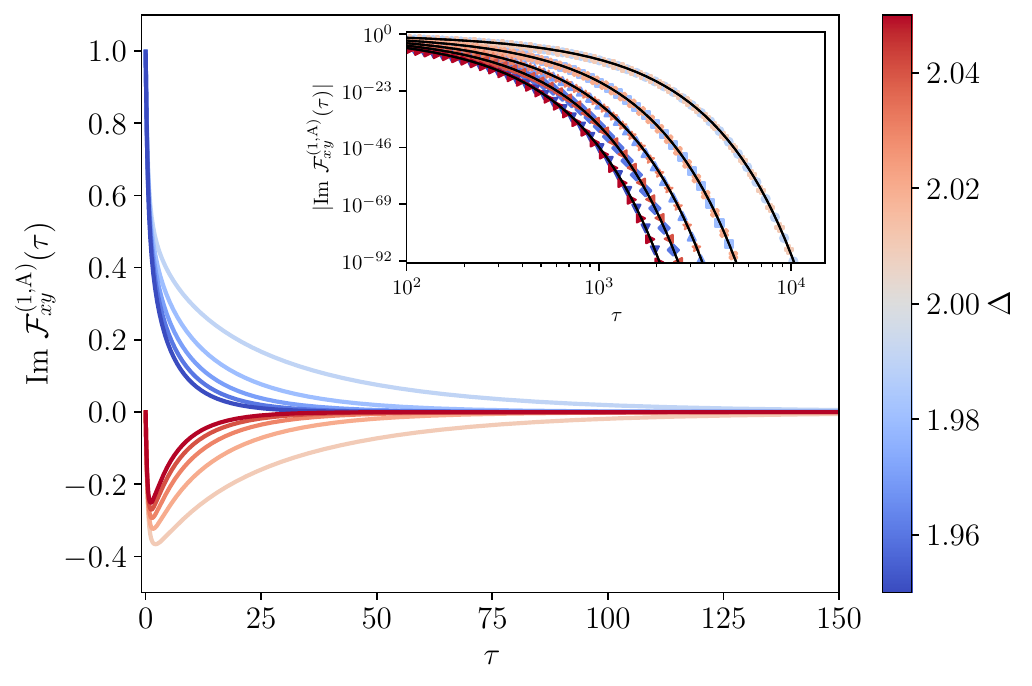}
  \qquad \quad  \includegraphics[width=0.4\linewidth]{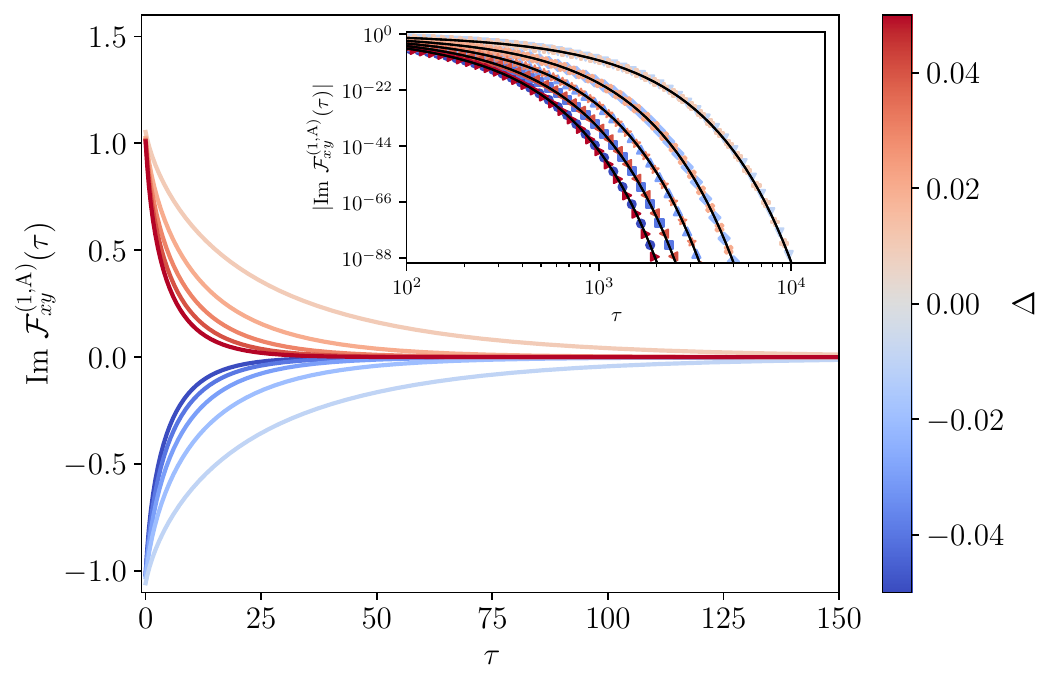}
    \caption{Numerical results for the pristine QWZ model, showing
$\text{Im}\,\mathcal{F}_{xy}^{(1,\mathrm{A})}(\mathbf{R})$ (upper panels) and
$\text{Im}\,\mathcal{F}_{xy}^{(1,\mathrm{A})}(\tau)$ (lower panels). The left-hand panels correspond to the transition from $C=1$ to $C=0$, whereas the right-hand panels show the transition from $C=-1$ to $C=1$, upon increasing $\Delta$. The insets demonstrate the agreement between the
numerical results and the analytical predictions for the asymptotic behaviour.}
    \label{fig:qwz_nonlocal}
\end{figure*}

We report in Fig.~\ref{fig:qwz_nonlocal} numerical results for the QWZ model; upper (lower) panels show the spatial (temporal) behaviour of the imaginary part of the position correlator.  Around $\Delta_c=0$, there is a transition from Chern number $C=-1$ to $C=1$ (right-hand panels), upon increasing $\Delta$. Meanwhile, for $\Delta_c=2$, there is a transition from $C=1$ to $C=0$ (left-hand panels). Here we show that in the asymptotic limit, neither the specific value of the topological invariant nor the magnitude of its jump across the transition matters: all real-space correlators decay as Eq.~\ref{eq:r_asymptotic}, where the Dirac fermion mass $m=\Delta-\Delta_c$ measures the distance from the critical point, while all imaginary-time correlators follow Eq.~\ref{eq:t_asymptotic}. Since the model has $C_2$ symmetry, $\mathcal{F}_{xy}^{(1)}(\mathbf{R})$ has a periodicity of 2 unit cells. 

\section*{References}
\bibliography{biblio}
\onecolumngrid 
\end{document}